\documentclass[onecolumn,english,aps,prl,amsmath,amssymb,superscriptaddress]{revtex4}
\usepackage[latin9]{inputenc}
\usepackage{color}
\usepackage{amstext}
\usepackage{amssymb}
\usepackage{graphicx}

\makeatletter
\@ifundefined{textcolor}{}
{%
 \definecolor{BLACK}{gray}{0}
 \definecolor{WHITE}{gray}{1}
 \definecolor{RED}{rgb}{1,0,0}
 \definecolor{GREEN}{rgb}{0,1,0}
 \definecolor{BLUE}{rgb}{0,0,1}
 \definecolor{CYAN}{cmyk}{1,0,0,0}
 \definecolor{MAGENTA}{cmyk}{0,1,0,0}
 \definecolor{YELLOW}{cmyk}{0,0,1,0}
}

\usepackage{color}

\newcommand{\beq}{\begin{eqnarray}}
\newcommand{\eeq}{\end{eqnarray}}
\newcommand{\ys}[1]{\textcolor{black}{#1}}
\makeatother
\usepackage{babel}

\begin{document}

\title{
Evolution of superconductivity and charge order in pressurized RbV$_3$Sb$_5$
}

\author{Feng Du}


\affiliation{Center for Correlated Matter and Department of Physics, Zhejiang University, Hangzhou 310058, China}
\affiliation  {Zhejiang Province Key Laboratory of Quantum Technology and Device, Department of Physics, Zhejiang University, Hangzhou 310058, China}

\author{Shuaishuai Luo}


\affiliation{Center for Correlated Matter and Department of Physics, Zhejiang University, Hangzhou 310058, China}
\affiliation  {Zhejiang Province Key Laboratory of Quantum Technology and Device, Department of Physics, Zhejiang University, Hangzhou 310058, China}

\author{Rui Li}

\affiliation{Center for Correlated Matter and Department of Physics, Zhejiang University, Hangzhou 310058, China}
\affiliation  {Zhejiang Province Key Laboratory of Quantum Technology and Device, Department of Physics, Zhejiang University, Hangzhou 310058, China}

\author{Brenden R. Ortiz}

\affiliation{Materials Department and California Nanosystems Institute, University of California Santa Barbara, Santa Barbara, CA, 93106, United States}

\author{Ye Chen}

\affiliation{Center for Correlated Matter and Department of Physics, Zhejiang University, Hangzhou 310058, China}
\affiliation  {Zhejiang Province Key Laboratory of Quantum Technology and Device, Department of Physics, Zhejiang University, Hangzhou 310058, China}

\author{Stephen D. Wilson}

\affiliation{Materials Department and California Nanosystems Institute, University of California Santa Barbara, Santa Barbara, CA, 93106, United States}

\author{Yu Song}

\email{yusong\_phys@zju.edu.cn}

\affiliation{Center for Correlated Matter and Department of Physics, Zhejiang University, Hangzhou 310058, China}
\affiliation  {Zhejiang Province Key Laboratory of Quantum Technology and Device, Department of Physics, Zhejiang University, Hangzhou 310058, China}

\author{Huiqiu Yuan}

\email{hqyuan@zju.edu.cn}

\selectlanguage{english}%

\affiliation{Center for Correlated Matter and Department of Physics, Zhejiang University, Hangzhou 310058, China}
\affiliation  {Zhejiang Province Key Laboratory of Quantum Technology and Device, Department of Physics, Zhejiang University, Hangzhou 310058, China}
\affiliation  {State Key Laboratory of Silicon Materials, Zhejiang University, Hangzhou 310058, China}

\begin{abstract}
The kagome metals $A$V$_3$Sb$_5$ ($A=$~K, Rb, Cs) under ambient pressure exhibit an unusual charge order, from which superconductivity emerges. 
In this work, by applying hydrostatic pressure using a liquid pressure medium and carrying out electrical resistance measurements for RbV$_3$Sb$_5$, we find the charge order becomes suppressed under a modest pressure $p_{\rm c}$ ($1.4<p_{\rm c}<1.6$~GPa), while the superconducting transition temperature $T_{\rm c}$ is maximized. $T_{\rm c}$ is then gradually weakened with further increase of pressure \ys{and reaches a minimum around 14.3~GPa,} before exhibiting another \ys{maximum} around 22.8~GPa, signifying the presence of a second superconducting dome. Distinct normal state resistance anomalies are found to be associated with the second superconducting dome, similar to KV$_3$Sb$_5$.
Our findings point to qualitatively similar temperature-pressure phase diagrams in KV$_3$Sb$_5$ and RbV$_3$Sb$_5$, \ys{and suggest a close link} between the second superconducting dome and the high-pressure resistance anomalies.  
\end{abstract}

\maketitle

\section{Introduction}

The kagome metals $A$V$_3$Sb$_5$ ($A=$~K, Rb, Cs) \cite{Ortiz2019,Ortiz2020,yin2021superconductivity,ortiz2020superconductivity} have drawn significant interest recently, with focus on (1) superconductivity with topological surfaces states \cite{Ortiz2020}, which raises the prospects of realizing topological superconductivity and Majorana zero modes \cite{wang2020proximityinduced,liang2021threedimensional}; (2) a large anomalous Hall effect in the absence of local moments \cite{Yang2020,yu2021concurrence,kenney2020absence}, which possibly results from a charge order that breaks time-reversal symmetry \cite{jiang2020discovery,Feng2021,setty2021electron,lin2021kagome}; (3) the nature of the superconducting pairing, with evidence for nodeless superconducting gaps \cite{duan2021nodeless,mu2021swave,xu2021multiband}, as well as indications for nematic superconductivity \cite{xiang2021twofold}; and (4) the sensitivity of superconductivity and charge order to pressure-tuning \cite{zhao2021nodal,Yu2021,chen2021double,Du2021,ChenCPL,Zhang2021,zhu2021doubledome,yin2021strainsensitive,wang2021competition}, likely related to the presence of competing instabilities on the kagome lattice \cite{Wang2013,Isakov2006,Guo2009,Kiesel2013,Wen2010}.

With the application of pressure, the charge order becomes quickly suppressed and superconductivity is enhanced, forming a superconducting dome with maximal $T_{\rm c}$ near the pressure at which charge order disappears \cite{zhao2021nodal,Yu2021,chen2021double,Du2021,ChenCPL,Zhang2021,zhu2021doubledome,wang2021competition}. While this superconducting dome can be understood to result from the competition between superconductivity and charge order, a second superconducting dome is found in $A$V$_3$Sb$_5$ at higher pressures \cite{zhao2021nodal,Du2021,ChenCPL,Zhang2021,zhu2021doubledome}, whose origin remains unclear. Factors that may be relevant for \ys{the} formation of the second superconducting dome include a pressure-induced Lifshitz transition \cite{ChenCPL}, magnetism suppressing superconductivity in the region between the two superconducting domes \cite{zhang2021firstprinciples}, reconstruction of the Sb bands due to the formation of interlayer Sb-Sb bonds \cite{tsirlin2021anisotropic}, and a distinct high-pressure phase revealed through resistivity anomalies \cite{Du2021}. 
While two-dome superconductivity is found in all $A$V$_3$Sb$_5$ materials via resistivity measurements \cite{zhu2021doubledome}, studies of the corresponding normal state from which superconductivity emerges has been limited \cite{Du2021,ChenCPL}. In particular, it remains unclear whether resistivity anomalies associated with \ys{the second superconducting dome} in KV$_3$Sb$_5$ \cite{Du2021} are also present in RbV$_3$Sb$_5$ and CsV$_3$Sb$_5$. 

In addition, single crystal diffraction revealed a highly anisotropic compression under hydrostatic pressure, with the $c$ lattice parameter reducing significantly more than the in-plane lattice parameters \cite{tsirlin2021anisotropic}, which suggests hydrostaticity of measurements under pressure may strongly affect experimental results and the determined phase diagrams. Therefore, to compare results on the $A$V$_3$Sb$_5$ series \ys{obtained} using different experimental techniques, it is important to reproduce the corresponding experimental hydrostaticity.

In this work, we systematically investigated the temperature-pressure phase diagram of RbV$_3$Sb$_5$ via measurements of electrical resistance, applying pressure using a liquid pressure medium.
It is found that the superconducting transition temperature $T_{\rm c}$ in RbV$_3$Sb$_5$ increases from $0.9$~K under ambient pressure to \ys{about} 4.5~K at 1.2~GPa, with the charge order disappearing above $p_{\rm c}$, \ys{which occurs between 1.4 and 1.6~GPa}. $T_{\rm c}$ is then gradually suppressed with increasing pressure \ys{and reaches a minimum of about 0.4~K around 14.3~GPa. With further increase of pressure, another $T_{\rm c}$ maximum of about 1.2~K is observed} around 22.8~GPa, indicating the presence of a second superconducting dome.
For pressures \ys{$\gtrsim22.8$~GPa, $T_{\rm c}$ in the second superconducting dome is suppressed with increasing pressure, and distinct features are detected in the normal state resistance, possibly related to a high-pressure phase. These behaviors are similar to pressurized KV$_3$Sb$_5$ \cite{Du2021}, suggesting that they are common in the $A$V$_3$Sb$_5$ series under high pressure, and are closely related to the suppression of superconductivity in the second dome.}

\section{Experimental Details}
Single crystals of RbV$_3$Sb$_5$ were grown using a self-flux method \cite{Ortiz2019,Ortiz2020}. Electrical resistance measurements under pressure were carried out using a diamond anvil cell (DAC), with silicon oil as the pressure medium, which allows for better hydrostaticity compared to solid pressure media. Single crystals of RbV$_3$Sb$_5$ were polished and cut into square pieces, and then loaded into a Be-Cu diamond anvil cell with a 400-$\mu$m-diameter culet. The ruby fluorescence method was used to determine the values of pressure inside the DAC \ys{at room temperature}, before and after the measurements. All electrical resistance measurements from room temperature down to 0.3~K were performed in a Teslatron-PT system with an Oxford $^3$He insert. For measurements with an applied magnetic field, the field was applied along the $c$-axis.

\section{Results}

\begin{figure*}
	\includegraphics[scale=0.6]{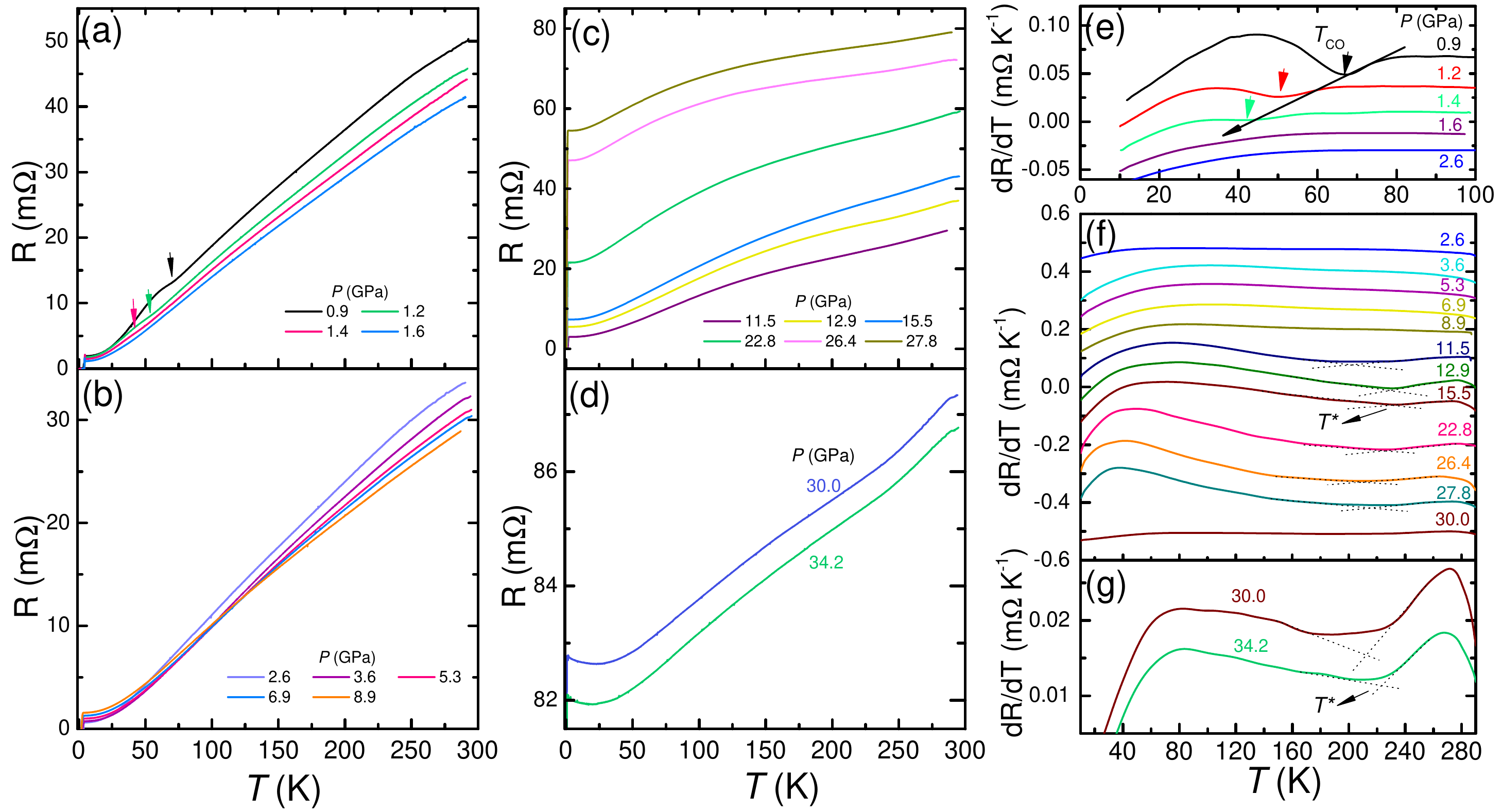} \protect\caption{(a)-(d): Electrical resistance $R(T)$ of RbV$_3$Sb$_5$ \ys{sample \#1} in various pressure ranges.
		(e)-(g): The corresponding d$R$/d$T$ curves, shifted vertically for clarity. $T_{\rm CO}$ is the onset temperature of the charge order, and $T^*$ corresponds to \ys{a minimum} in d$R$/d$T$. 
	}
	\label{Fig_R}
\end{figure*}

Measurements of the electrical resistance $R(T)$ \ys{for sample \#1} with pressures from 0.9~GPa up to 34.2~GPa are shown in Figs.~\ref{Fig_R}(a)-(d), with the corresponding d$R$/d$T$ curves shown in Figs.~\ref{Fig_R}(e)-(g). Examining the behavior of $R(T)$ in the normal state, clear anomalies associated with the charge ordering temperature $T_{\rm CO}$ can be seen up to $1.4$~GPa, with $T_{\rm CO}$ decreasing with increasing pressure [Fig.~\ref{Fig_R}(a)]. The values of $T_{\rm CO}$ can be determined from the anomaly in the d$R$/d$T$ curves, as shown in Fig.~\ref{Fig_R}(e), which becomes indiscernible at $1.6$~GPa. These results suggest the charge order disappears at a critical pressure $p_{\rm c}$ between 1.4 and 1.6~GPa. $p_{\rm c}$ in RbV$_3$Sb$_5$ is larger than that in KV$_3$Sb$_5$ ($\approx0.5$~GPa) \cite{Du2021}, consistent with the idea that larger alkaline metal ions exert negative chemical pressure, although it should be noted that additional effects may also be at play, as suggested by the complex evolution of $T_{\rm c}$ with pressure in CsV$_3$Sb$_5$ \ys{and RbV$_3$Sb$_5$} for $p<p_{\rm c}$ \cite{chen2021double,Yu2021,wang2021competition}.

In the pressure range 2.6\ys{~GPa} to 8.9~GPa, $R(T)$ in the normal state is largely unchanged with increasing pressure [Fig.~\ref{Fig_R}(b)]. In contrast, in the pressure range 11.5\ys{~GPa} to 27.8~GPa, the values of $R(T)$ increase with increasing pressure, with the increase particularly prominent for $p\gtrsim22.8$~GPa [Fig.~\ref{Fig_R}(c)]. This difference between the results in Figs.~\ref{Fig_R}(b) and (c) is largely due to the residual resistance $R_0$, which is the value of $R(T)$ just before the onset of superconductivity. On the other hand, the variation of $R(T)$ with temperature, which can be captured by $\delta R=R(300~{\rm K})-R_0$, stays nearly unchanged from 2.6\ys{~GPa} to 22.8~GPa, but starts to decrease dramatically when $p\gtrsim22.8$~GPa. \ys{Clear anomalies in d$R$/d$T$, with a \ys{minimum} at $T^*\sim220$~K [Figs.~\ref{Fig_R}(f) and (g)], are detected for $p\gtrsim11.5$~GPa. It should be noted that since d$R$/d$T$ exhibits multiple anomalies around $T^*$, with possible additional features above room temperature, $T^*$ should be considered as a characteristic temperature rather than a transition temperature.} 

For even higher pressures of 30.0\ys{~GPa} and 34.2~GPa, the normal state $R(T)$ becomes qualitatively different: compared to lower pressures, the values of $\delta R$ are much smaller, $R_0$ becomes much larger [Fig.~\ref{Fig_R}(d)], \ys{and a subtle upturn in $R(T)$ is observed for $T\lesssim20$~K}. \ys{In addition, the d$R$/d$T$ anomaly associated with $T^*$ becomes more dominant for $p\gtrsim30.0$~GPa, due to the dramatic decrease of $\delta R$ (and thus the magnitude of d$R$/d$T$) at these pressures.} These behaviors are similar to KV$_3$Sb$_5$ at high pressures \cite{Du2021}, and can be \ys{collectively} attributed to the formation of a high-pressure phase. 

Since the high-pressure phase exhibits a small $\delta R$ and a large $R_0$, the observation that $\delta R$ begins to drop while $R_0$ begins to increase for $p\gtrsim22.8$~GPa suggests that the high-pressure phase \ys{onsets around} 22.8~GPa. 
\ys{On the other hand, signatures of $T^{*}$ already appears for 11.5~GPa, a significantly lower pressure compared to when $\delta R$ begins to drop and $R_0$ begins to increase, suggesting that is not unique to the high-pressure phase.} More work is needed to definitively pin down the phase boundary of the high-pressure phase in the temperature-pressure phase diagram \ys{and elucidate the origin of the resistance anomalies around $T^*$}.

\begin{figure}
	\includegraphics[scale=0.5]{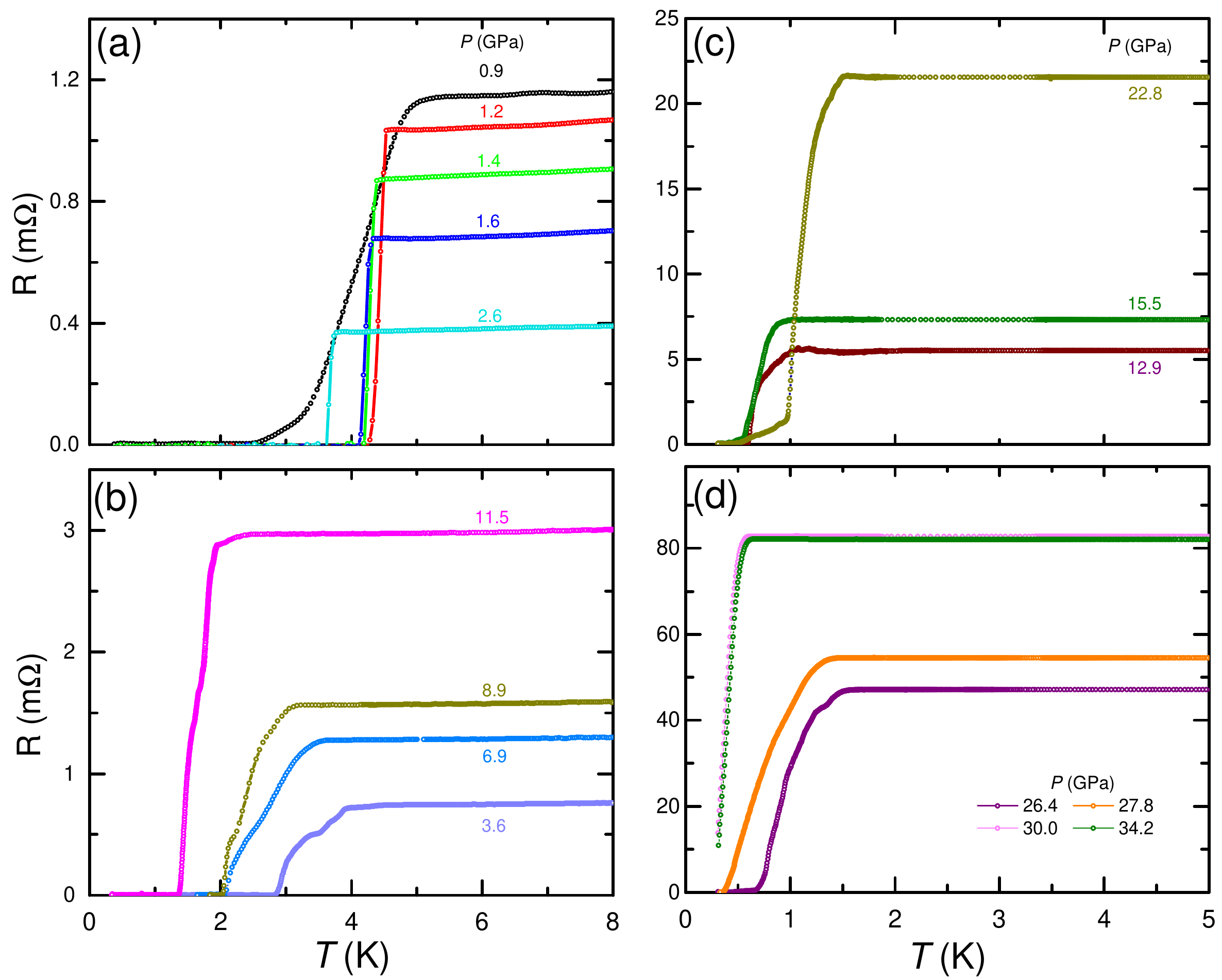} \protect\caption{Zoomed-in electrical resistance $R(T)$ of RbV$_3$Sb$_5$ \ys{sample \#1} under various pressures
	to show the evolution of superconductivity with pressure.}
	\label{Fig_Tc}
\end{figure}

Fig.~\ref{Fig_Tc} zooms into low temperatures to examine the evolution of superconductivity in RbV$_3$Sb$_5$ \ys{sample \#1}, as a function of applied pressure. For pressures from 0.9\ys{~GPa} to 2.6~GPa [Fig.~\ref{Fig_Tc}(a)], $T_{\rm c}$ is bunched around 4~K, significantly higher than $T_{\rm c}<1$~K under ambient pressure \cite{yin2021superconductivity}. A prominent feature is that the superconducting transition is significantly broader for 0.9~GPa, compared to those from 1.2 to 2.6~GPa, which suggests that a broad superconducting transition is correlated with the presence of a strong charge order. It should be noted that while signatures of the charge order is also detected for 1.2\ys{~GPa} and 1.4~GPa, the corresponding dip in d$R$/d$T$ is significantly weakened compared to 0.9~GPa, which indicates that the magnitude of the charge order parameter \ys{becomes} significantly reduced. A similar sharpening of the superconducting transition with weakening of the charge order upon applying pressure is also observed in KV$_3$Sb$_5$ \cite{Du2021}, suggesting that it is a common feature of the $A$V$_3$Sb$_5$ series. 

\begin{figure}
	\includegraphics[scale=0.5]{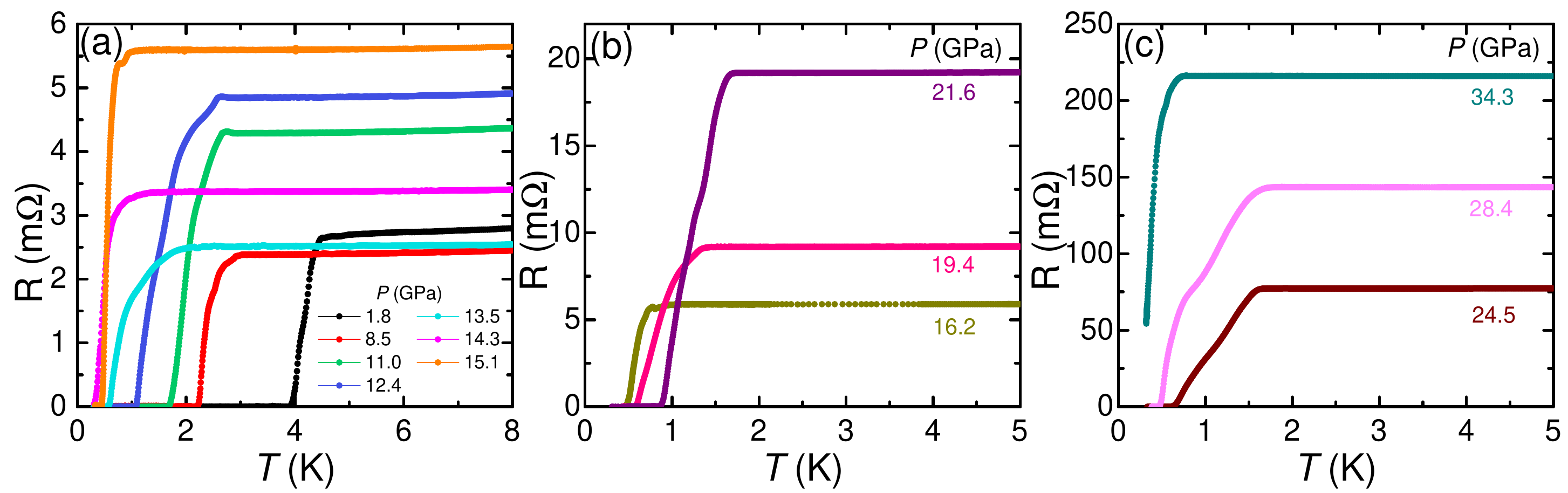} \protect\caption{\ys{Electrical resistance $R(T)$ of RbV$_3$Sb$_5$ sample \#2 near the superconducting transition under various pressures,
	showing the evolution of superconductivity with pressure.}}
	\label{Fig_Tc_2}
\end{figure}

One possible origin for this behavior is that d$T_{\rm c}$/d$p$ is much larger in the presence of a strong charge order, compared to when the charge order is significantly weakened or suppressed. If there is a distribution of internal stress $\Delta p$ in the sample (\ys{and assume that it does} not change too much with applied pressure), then the superconducting transition width $\Delta T_{\rm c}\approx\frac{{\rm d}T_{\rm c}}{{\rm d}p}\Delta p$, would be large when d$T_{\rm c}$/d$p$ is large. 

Alternatively, domain boundaries of a period-2 charge order could exhibit enhanced filamentary superconductivity, leading to a broad superconducting transition in $R(T)$. Such a picture was invoked to account for enhanced superconductivity seen in the period-2 charge order phase of Ba$_{1-x}$Sr$_x$Ni$_2$As$_2$, while such an effect is absent for the period-3 charge order phase in the same material series \cite{Lee2021}. Given the charge order in $A$V$_3$Sb$_5$ is period-2 in the $ab$-plane \cite{Ortiz2020}, such a mechanism may also contribute to its broadened superconducting transition under ambient and low pressures, where the charge order is substantial. 

In the pressure range 3.6\ys{~GPa} to 11.5~GPa [Fig.~\ref{Fig_Tc}(b)], $T_{\rm c}$ gradually reduces with increasing pressure, which in combination with results in Fig.~\ref{Fig_Tc}(a), evidence a superconducting dome with maximal $T_{\rm c}$ around $p_{\rm c}$. Such a behavior was previously reported for the $A$V$_3$Sb$_5$ materials \cite{zhao2021nodal,Yu2021,chen2021double,zhu2021doubledome,Du2021}, and likely results from a competition between charge order and superconductivity, without a prominent role of quantum criticality \cite{Du2021}.

Upon further increasing pressure, $T_{\rm c}$ further drops for 12.9\ys{~GPa} and 15.5~GPa, but becomes enhanced again at 22.8~GPa [Fig.~\ref{Fig_Tc}(c)]. This suggests the presence of a minimum in $T_{\rm c}$ between \ys{12.9~GPa} and 22.8~GPa, and signifies the appearance of a second superconducting dome. For pressures in the range 26.4\ys{~GPa} to 34.2~GPa [Fig.~\ref{Fig_Tc}(d)], $T_{\rm c}$ decreases with increasing pressure, indicating that the maximal $T_{\rm c}$ of the second superconducting dome appears at $p<26.4$~GPa. Moreover, the superconducting transition becomes sharpened for 30.0 and 34.2~GPa, possibly related to the high-pressure phase, with clear signatures in the normal state. It is interesting to note that the suppression of superconductivity in RbV$_3$Sb$_5$
is less significant compared to KV$_3$Sb$_5$, as the high-pressure phase appears. Compared to superconducting transitions for $p\leq2.6$~GPa, those for $p\geq3.6$~GPa are generally broader and less smooth, possibly related to a larger distribution of stress or pressure-induced defects in the crystal.


\ys{To clarify the evolution of $T_{\rm c}$ near the boundary of the two superconducting domes, we measured $R(T)$ near $T_{\rm c}$ for RbV$_3$Sb$_5$ sample \#2, with results shown in Fig.~\ref{Fig_Tc_2}. Upon increasing pressure from 1.8~GPa up to 14.3~GPa [Fig.~\ref{Fig_Tc_2}(a)], $T_{\rm c}$ monotonically decreases with increasing pressure, with the value of $T_{\rm c}$ (determined through the midpoint of the superconducting transition) reaching a minimum of about 0.4~K at 14.3~GPa. With further increase of pressure, $T_{\rm c }$ becomes slightly enhanced at 15.1~GPa, and furthers increases with increasing pressure up to 21.6~GPa [Fig.~\ref{Fig_Tc_2}]. Upon further increase of pressure, $T_{\rm c}$ again reduces with increasing pressure [Fig.~\ref{Fig_Tc_2}]. These measurements on sample \#2 are consistent with those for sample \#1 [Fig.~\ref{Fig_Tc}] in establishing the presence of two superconducting domes, and further clarifies that the minimum in $T_{\rm c}$ between the two domes occurs around 14.3~GPa.}

\begin{figure}
	\includegraphics[scale=0.5]{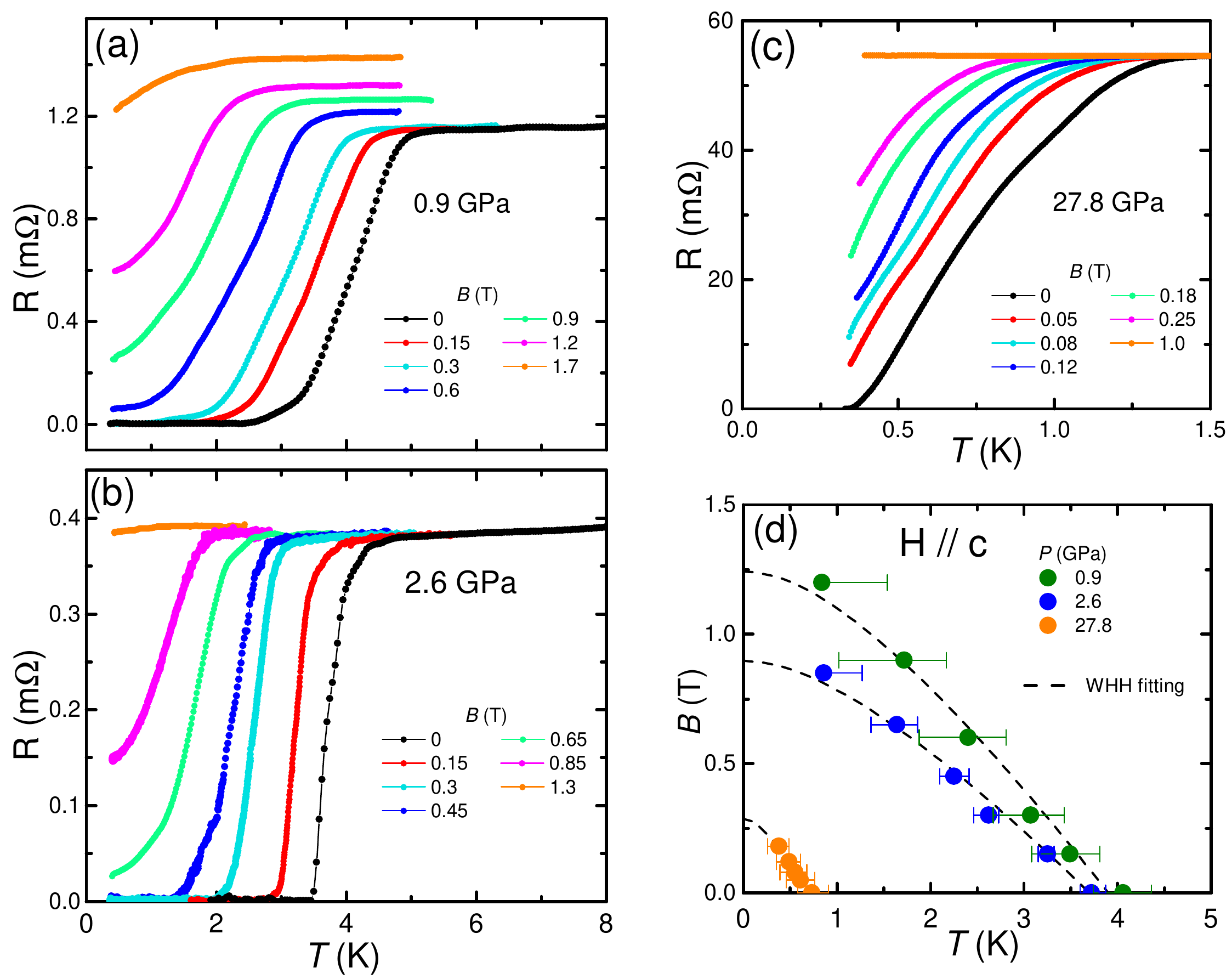} \protect\caption{Low-temperature electrical resistance $R(T)$ of RbV$_3$Sb$_5$ \ys{sample \#1} under various $c$-axis magnetic fields, for applied pressures (a) 0.9~GPa, (b) 2.6~GPa, and (c) 27.8~GPa. (d) The upper critical fields of RbV$_3$Sb$_5$, as a function of temperature under pressures of 0.9, 2.6, and 27.8~GPa. Fits to the WHH model are shown as dashed lines.}
	\label{Fig_mag}
\end{figure}

To probe the superconducting state under various pressures, we chose three representative pressures and measured the resistance \ys{for sample \#1} under an applied magnetic field along the $c$-axis, with results shown in Fig.~\ref{Fig_mag}. The three pressures correspond to states with coexisting charge order and superconductivity [0.9~GPa, Fig.~\ref{Fig_mag}(a)], without charge order and inside the first superconducting dome [2.6~GPa, Fig.~\ref{Fig_mag}(b)], and inside the second superconducting dome [27.8~GPa, Fig.~\ref{Fig_mag}(c)]. As can be seen, superconductivity is gradually suppressed with applied field in all cases, which allows us to extract the upper critical field $\mu_0 H_{\rm{c2}}(T)$ for various pressures.

The upper critical fields $\mu_0 H_{\rm{c2}}(T)$ are determined as when $R(T)$ drops to $R_0/2$, and are summarized in Fig.~\ref{Fig_mag}(d). $\mu_0 H_{\rm{c2}}(T)$ could be fit with the Werthamer-Helfand-Hohenberg (WHH) model \cite{WHH} for the three pressures, with fits shown as \ys{dashed} lines in Fig.~\ref{Fig_mag}(d). Previously, it was shown in KV$_3$Sb$_5$ that a much larger $\mu_0 H_{\rm{c2}}(T=0)$ (for field along the $c$-axis) is observed for the second superconducting dome, even when the $T_{\rm c}$ is lower compared to the first superconducting dome \cite{Du2021}. For RbV$_3$Sb$_5$, the values of $T_{\rm c}^2/\mu_0 H_{\rm{c2}}(T=0)$ are respectively found to be 13.3, 15.2 and 1.9 (K$^2$/T), for 0.9\ys{~GPa}, 2.6\ys{~GPa} and 27.8~GPa. The significantly smaller ratio at 27.8~GPa points to $\mu_0 H_{\rm{c2}}(T=0)$ also becoming  enhanced relative to $T_{\rm c}$ in the second superconducting dome, similar to KV$_3$Sb$_5$.

\begin{figure}
	\includegraphics[scale=0.5]{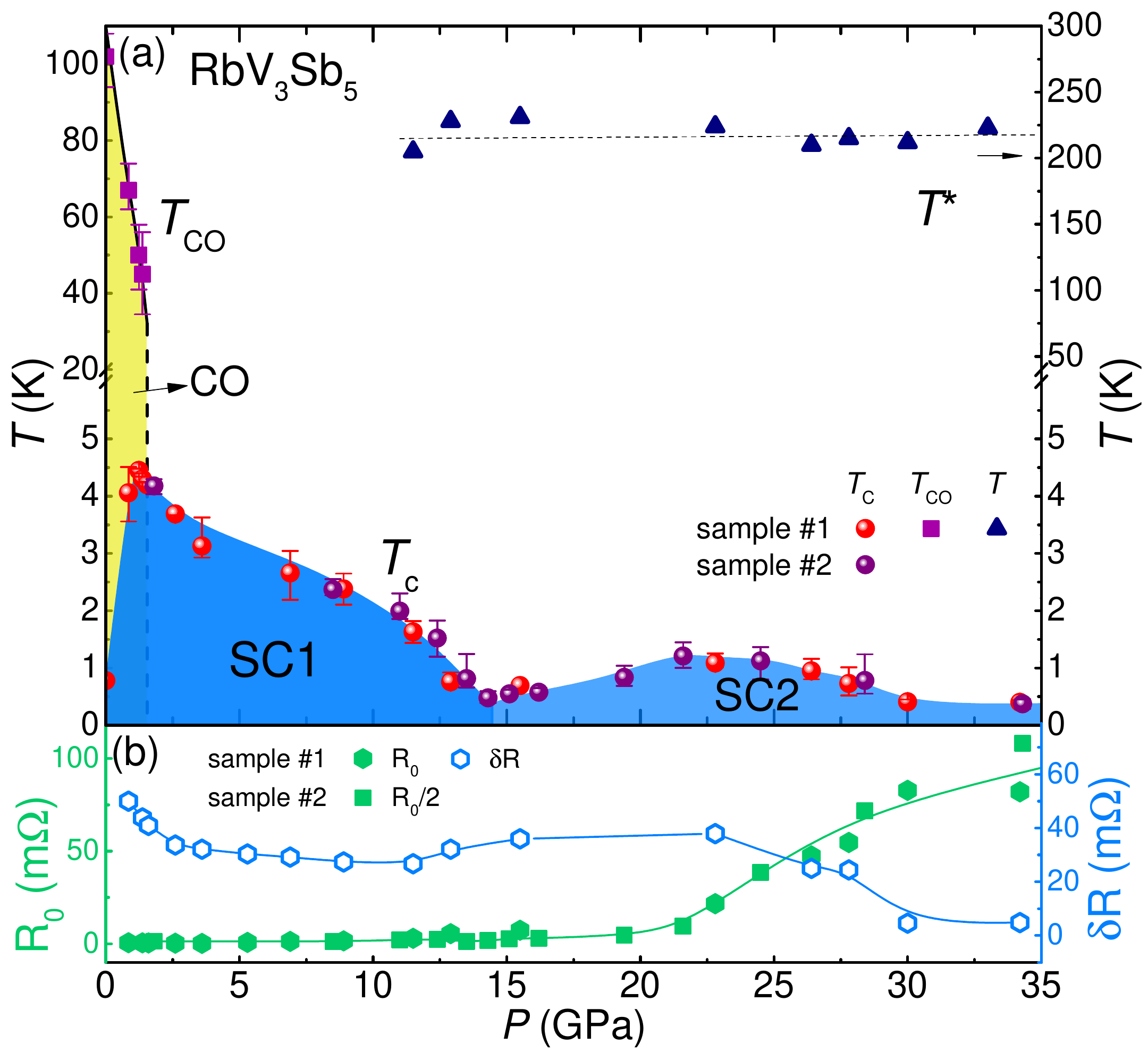} \protect\caption{(a) The temperature-pressure phase diagram of RbV$_3$Sb$_5$. Two superconducting regions (SC1 and SC2) and a charge-ordered phase (CO) are identified. Points under ambient pressure are from Ref.~\cite{yin2021superconductivity}. $T_{\rm CO}$ is the onset temperature of the charge order, $T_{\rm c}$ is the superconducting transition temperature, and $T^*$ is a characteristic temperature associated \ys{with a minimum in d$R$/d$T$}. \ys{The errors bars for $T_{\rm CO}$ are determined from the half width at half minimum of the dip in d$R$/d$T$. The errors bars for $T_{\rm c}$ characterize the width of the superconducting transition, with values corresponding to 20\% and 80\% of $R_0$, where $R_0$ is the resistance just before the onset of superconductivity. In cases where $R(T)$ does not drop to 20\% of $R_0$ at the lowest measured temperature, $R(T)$ is extrapolated to obtain an estimate.} (b) Pressure dependence of the residual resistance $R_0$ and the electrical resistance change $\delta R=R(300{\rm~K})-R_0$. \ys{$R_0$ for sample \#2 is divided by 2, and plotted together with $R_0$ for sample \#1. $\delta R$ was only measured for sample \#1.} The solid lines are guides-to-the-eye.}
	\label{Fig_phase_diagram}
\end{figure}

The phase diagram obtained from electrical resistance measurements under pressure is shown in Fig.~\ref{Fig_phase_diagram}(a). $T_{\rm c}$ is determined as when $R(T)$ drops to $R_0/2$, $T^*$ from \ys{a minimum} in d$R$/d$T$ [Fig.~\ref{Fig_R}(g)], and $T_{\rm CO}$ from the dip in d$R$/d$T$ [Fig.~\ref{Fig_R}(e)]. In KV$_3$Sb$_5$, it was shown that the decrease of $T_{\rm c}$ above $p_{\rm c}$ is much more gradual, compared to the enhancement of $T_{\rm c}$ below $p_{\rm c}$, resulting in the first superconducting dome being highly asymmetric \cite{Du2021}. In the case of RbV$_3$Sb$_5$, it can be seen that the same qualitative behavior is observed.  



Various regimes in the phase diagram can be identified through \ys{the presence of a resistivity anomaly at $T^{*}$ [Fig.~\ref{Fig_phase_diagram}(a)] and} signatures in $\delta R$ and $R_0$ [Fig.~\ref{Fig_phase_diagram}(b)]. \ys{The anomaly in d$R$/d$T$ associated with $T^{*}$ first appears at 11.5~GPa, just before superconductivity crossovers from the first dome into the second.} $R_0$ increases slowly with increasing pressure up to $\approx22.8$~GPa, above which it increases at a much faster rate up to $\approx30$~GPa, and finally plateaus for $p\gtrsim30$~GPa. \ys{$\delta R$ exhibits a gradual decrease with increasing pressure up to 1.4~GPa ($\approx p_{\rm c}$), then evolves slowly up to $\approx22.8$~GPa, before dropping precipitously for $p\gtrsim22.8$~GPa. Since $T_{\rm c}$ in the second superconducting dome begins to decrease with increasing pressure for $p\gtrsim22.8$~GPa, a large $R_0$ and a small $\delta R$, which are signatures of a possible high-pressure phase, appear to be correlated with suppression of $T_{\rm c}$ in the second superconducting dome. A similar large drop of $\delta R$ and an increase in $R_0$ are also observed in KV$_3$Sb$_5$, and are also associated with the suppression of $T_{\rm c}$ in the second superconducting dome \cite{Du2021}.} 

\section{Discussion and Conclusion}
\begin{figure}
	\includegraphics[scale=0.5]{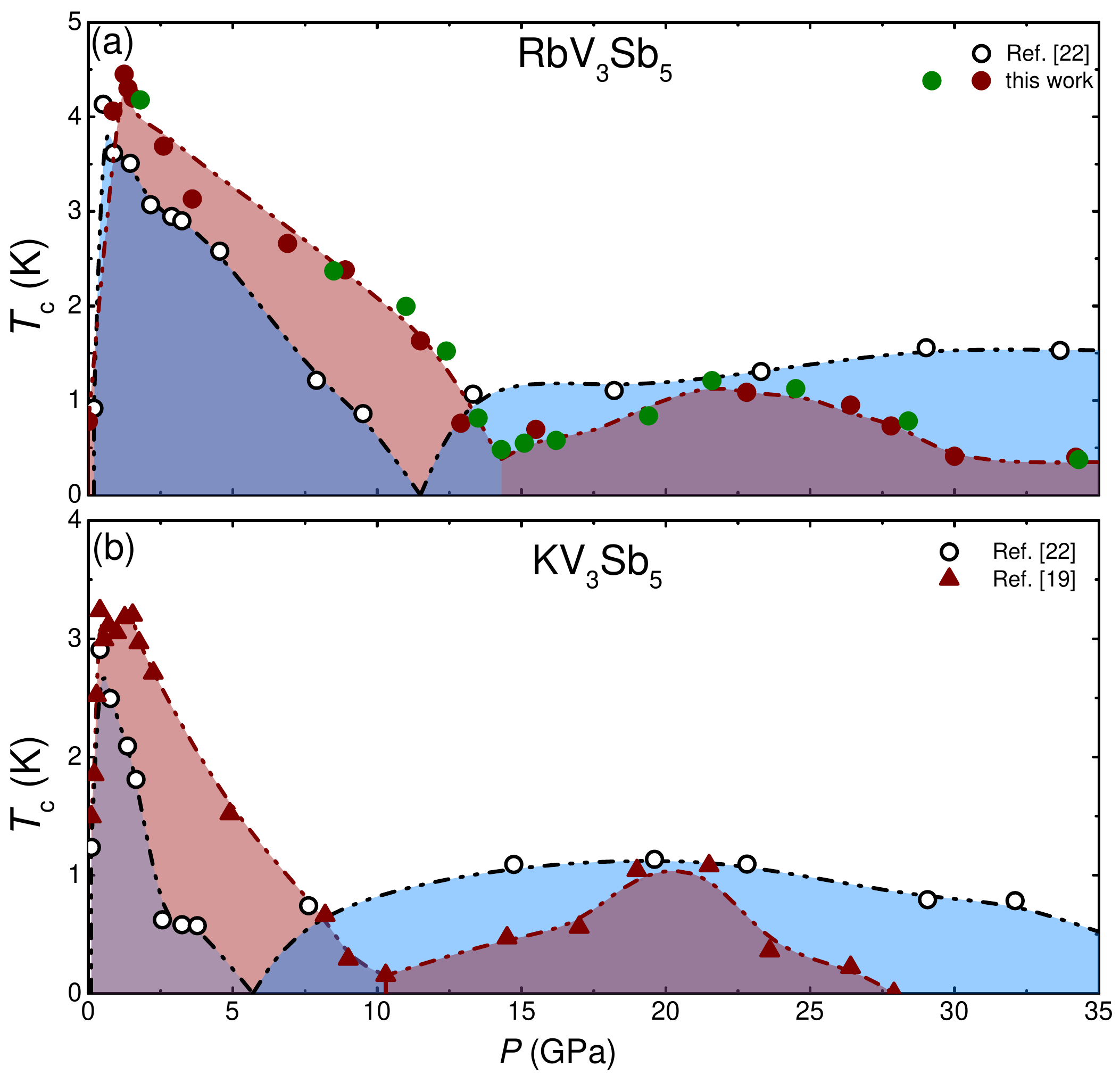} \protect\caption{(a) Comparison between superconducting phase diagrams of RbV$_3$Sb$_5$ from this work and Ref.~\cite{zhu2021doubledome}. \ys{The maroon symbols are from RbV$_3$Sb$_5$ sample \#1, and the green symbols are from sample \#2.} (b) Comparison between superconducting phase diagrams of KV$_3$Sb$_5$ from Refs.~\cite{zhu2021doubledome,Du2021}. $T_{\rm c}$ in this work and Ref.~\cite{Du2021} is defined as when the resistance(resistivity) drops to $R_0/2$($\rho_0/2$), whereas it is determined from $0.9\rho_0$ in Ref.~\cite{zhu2021doubledome}. The dashed lines are guides-to-the-eye.}
	\label{Fig_compare}
\end{figure}

Our observation of two-dome superconductivity is similar to previous works on $A$V$_3$Sb$_5$ \cite{zhao2021nodal,Du2021,ChenCPL,Zhang2021,zhu2021doubledome}, and suggests that such a feature is qualitatively robust, both with regards to variation of the alkaline metal and experimental hydrostaticity. On the other hand, the additional modulation of $T_{\rm c}$ for $p<p_{\rm c}$ in CsV$_3$Sb$_5$ \ys{and RbV$_3$Sb$_5$} \cite{Yu2021,chen2021double,wang2021competition} is not observed in KV$_3$Sb$_5$ \cite{Du2021}. \ys{While the present data on RbV$_3$Sb$_5$ suggests a single peak in $T_{\rm c}$ around $p_{\rm c}$, the density of probed pressures is insufficient to distinguish whether additional subtle modulations of $T_{\rm c}$ exist for $p<p_{\rm c}$ \cite{wang2021competition}.} 

Despite the qualitative agreement on the presence of two-dome superconductivity in the literature, directly comparing phase diagrams for RbV$_3$Sb$_5$ from this work and Ref.~\cite{zhu2021doubledome} [Fig.~\ref{Fig_compare}(a)], and \ys{for} KV$_3$Sb$_5$ from Refs.~\cite{Du2021,zhu2021doubledome} [Fig.~\ref{Fig_compare}(b)], reveal clear differences. The differences likely originate from hydrostaticity of the experiments, which either used silicon oil as a liquid pressure medium (\ys{solid} symbols, this work and Ref.~\cite{Du2021}), or cubic boron nitride as a solid pressure medium (\ys{open} symbols, Ref.~\cite{zhu2021doubledome}), with the former typically associated with better hydrostatcity. As can be seen, in the phase diagrams of both RbV$_3$Sb$_5$ and KV$_3$Sb$_5$ obtained using a liquid pressure medium, a larger pressure is required to suppress the first superconducting dome (and induce the onset of the second dome), compared to experiments using a solid pressure medium. On the other hand, superconductivity in the second dome appears more easily suppressed using a liquid medium, while it is more persistent when using a solid medium. While the origin for these systematic differences remain unclear, a key factor is the strongly anisotropic compression revealed under pressure, with the $c$-axis lattice parameter reduced by a staggering $\sim20$\% under 20~GPa in CsV$_3$Sb$_5$ \cite{tsirlin2021anisotropic}. Since typical studies on layered materials in a DAC applies force along the $c$-axis, nonideal hydrostaticity could lead to larger-than-nominal compressive pressure on the $ab$-plane. This may contribute to the smaller nominal pressures in experiments using a solid pressure medium, \ys{which are} needed to access, for example, the dip in $T_{\rm c}$ between the two superconducting domes.     
These observations point to the importance of hydrostaticity \ys{for experimentally determined phase diagrams}, which should be taken into \ys{consideration} when comparing experimental results \ys{on $A$V$_3$Sb$_5$ obtained under high-pressures}.


Comparing the present results on RbV$_3$Sb$_5$ with previous work on KV$_3$Sb$_5$ \cite{Du2021} [\ys{solid} symbols in Figs.~\ref{Fig_compare}(a) and (b)], it can be seen that a higher pressure is required to fully suppress the first superconducting dome in RbV$_3$Sb$_5$, consistent with the notion that a larger atomic radius corresponds to negative chemical pressure. More importantly, clear changes in the normal state transport ($R_0$ and $\delta R$), \ys{which can be attributed to a high-pressure phase,}
are seen under pressures associated with the \ys{suppression of $T_{\rm c}$ in the} second superconducting dome, for both RbV$_3$Sb$_5$ and KV$_3$Sb$_5$ \cite{Du2021}.  
\ys{With increasing pressure, the d$R$/d$T$ anomaly near $T^{*}$ first appears at the edge of the first superconducting dome, and persists across the second superconducting dome. 
Previous work on KV$_3$Sb$_5$ \cite{Du2021} also associated the resistance anomalies around $T^{*}$ with the high-pressure phase and the suppression of $T_{\rm c}$ in the second superconducting dome. Examination of the published data reveals that these anomalies in KV$_3$Sb$_5$ also extend to lower pressures, suggesting that they are not unique to the high-pressure phase, similar to the present results on RbV$_3$Sb$_5$.}

It is interesting to note that for pressures associated with the second superconducting dome in CsV$_3$Sb$_5$, Raman spectroscopy detected distinct signatures in the phonon spectrum \cite{ChenCPL}, and first-principles calculations found a structural instability \cite{zhang2021firstprinciples} and a reconstruction of the Fermi surface \cite{tsirlin2021anisotropic}, \ys{suggesting that the normal state from which superconductivity emerges in the second dome of CsV$_3$Sb$_5$ is distinct from that at lower pressures, similar to the present findings for RbV$_3$Sb$_5$.} Further work is needed to understand the nature of high-pressure phases in $A$V$_3$Sb$_5$, as well as clarify how they relate to the second superconducting dome.

In conclusion, we have systematically studied the electrical resistance of RbV$_3$Sb$_5$ by applying hydrostatic pressure via a liquid pressure medium, revealing the presence of two superconducting domes and \ys{distinct features in the normal state transport associated with the second dome}, similar to KV$_3$Sb$_5$. In both systems, the shape of the first superconducting dome is highly asymmetric, with maximal $T_{\rm c}$ near the border of the ambient pressure charge order. \ys{With increasing pressure, anomalies in d$R$/d$T$ near $T^*\sim 220$~K appears at the edge of the first superconducting dome and persists across the second dome. The suppression of superconductivity in the second superconducting dome is associated with significant changes in the normal state transport, which can be attributed to a high-pressure phase.} The commonality of \ys{a second superconducting dome at high pressures and behaviors in the normal state resistance} for both RbV$_3$Sb$_5$ and KV$_3$Sb$_5$ suggests a link between \ys{these} phenomena.

\section*{Acknowledgments}

This work was supported by the National Key R\&D Program of China (No. 2017YFA0303100, No. 2016YFA0300202), the Key R\&D Program of Zhejiang Province, China (2021C01002), the National Natural Science Foundation of China (No. 11974306 and No. 12034017), and the Fundamental Research Funds for the Central Universities of China.  S.D.W. and B.R.O. gratefully acknowledge support via the UC Santa Barbara NSF Quantum Foundry funded via the Q-AMASE-i program under award DMR-1906325.  B.R.O. also acknowledges support from the California NanoSystems Institute through the Elings fellowship program.
\bibliographystyle{iopart-num.bst}
\bibliography{bibfile}

\providecommand{\newblock}{}
\begin{thebibliography}{10}
\expandafter\ifx\csname url\endcsname\relax
  \def\url#1{{\tt #1}}\fi
\expandafter\ifx\csname urlprefix\endcsname\relax\def\urlprefix{URL }\fi
\providecommand{\eprint}[2][]{\url{#2}}

\bibitem{Ortiz2019}
Ortiz B~R, Gomes L~C, Morey J~R, Winiarski M, Bordelon M, Mangum J~S, Oswald
  I~W~H, Rodriguez-Rivera J~A, Neilson J~R, Wilson S~D, Ertekin E, McQueen T~M
  and Toberer E~S 2019 {\em Physical Review Materials\/} {\bf 3}
  \urlprefix\url{https://doi.org/10.1103/physrevmaterials.3.094407}

\bibitem{Ortiz2020}
Ortiz B~R, Teicher S~M, Hu Y, Zuo J~L, Sarte P~M, Schueller E~C, Abeykoon A~M,
  Krogstad M~J, Rosenkranz S, Osborn R, Seshadri R, Balents L, He J and Wilson
  S~D 2020 {\em Physical Review Letters\/} {\bf 125}
  \urlprefix\url{https://doi.org/10.1103/physrevlett.125.247002}

\bibitem{yin2021superconductivity}
Yin Q, Tu Z, Gong C, Fu Y, Yan S and Lei H 2021 {\em Chinese Physics Letters\/}
  {\bf 38} 037403 \urlprefix\url{https://doi.org/10.1088/0256-307x/38/3/037403}

\bibitem{ortiz2020superconductivity}
Ortiz B~R, Sarte P~M, Kenney E~M, Graf M~J, Teicher S~M~L, Seshadri R and
  Wilson S~D 2021 {\em Physical Review Materials\/} {\bf 5}
  \urlprefix\url{https://doi.org/10.1103/physrevmaterials.5.034801}

\bibitem{wang2020proximityinduced}
Wang Y, Yang S, Sivakumar P~K, Ortiz B~R, Teicher S~M~L, Wu H, Srivastava A~K,
  Garg C, Liu D, Parkin S~S~P, Toberer E~S, McQueen T, Wilson S~D and Ali M~N
  2020  (\textit{Preprint} \eprint{arXiv:2012.05898})

\bibitem{liang2021threedimensional}
Liang Z, Hou X, Zhang F, Ma W, Wu P, Zhang Z, Yu F, Ying J~J, Jiang K, Shan L,
  Wang Z and Chen X~H 2021 {\em Physical Review X\/} {\bf 11}
  \urlprefix\url{https://doi.org/10.1103/physrevx.11.031026}

\bibitem{Yang2020}
Yang S~Y, Wang Y, Ortiz B~R, Liu D, Gayles J, Derunova E, Gonzalez-Hernandez R,
  {\v{S}}mejkal L, Chen Y, Parkin S~S~P, Wilson S~D, Toberer E~S, McQueen T and
  Ali M~N 2020 {\em Science Advances\/} {\bf 6} eabb6003
  \urlprefix\url{https://doi.org/10.1126/sciadv.abb6003}

\bibitem{yu2021concurrence}
Yu F~H, Wu T, Wang Z~Y, Lei B, Zhuo W~Z, Ying J~J and Chen X~H 2021 {\em Phys.
  Rev. B\/} {\bf 104}(4) L041103
  \urlprefix\url{https://link.aps.org/doi/10.1103/PhysRevB.104.L041103}

\bibitem{kenney2020absence}
Kenney E~M, Ortiz B~R, Wang C, Wilson S~D and Graf M~J 2021 {\em Journal of
  Physics: Condensed Matter\/} {\bf 33} 235801
  \urlprefix\url{https://doi.org/10.1088/1361-648x/abe8f9}

\bibitem{jiang2020discovery}
Jiang Y~X, Yin J~X, Denner M~M, Shumiya N, Ortiz B~R, Xu G, Guguchia Z, He J,
  Hossain M~S, Liu X, Ruff J, Kautzsch L, Zhang S~S, Chang G, Belopolski I,
  Zhang Q, Cochran T~A, Multer D, Litskevich M, Cheng Z~J, Yang X~P, Wang Z,
  Thomale R, Neupert T, Wilson S~D and Hasan M~Z 2021 {\em Nature Materials\/}
  \urlprefix\url{https://doi.org/10.1038/s41563-021-01034-y}

\bibitem{Feng2021}
Feng X, Jiang K, Wang Z and Hu J 2021 {\em Science Bulletin\/}
  \urlprefix\url{https://doi.org/10.1016/j.scib.2021.04.043}

\bibitem{setty2021electron}
Setty C, Hu H, Chen L and Si Q 2021 {\em 2105.15204\/}
  \urlprefix\url{https://arxiv.org/abs/2105.15204}

\bibitem{lin2021kagome}
Lin Y~P and Nandkishore R~M 2021 {\em arXiv:2107.09050\/}
  \urlprefix\url{https://arxiv.org/abs/2107.09050}

\bibitem{duan2021nodeless}
Duan W, Nie Z, Luo S, Yu F, Ortiz B~R, Yin L, Su H, Du F, Wang A, Chen Y, Lu X,
  Ying J, Wilson S~D, Chen X, Song Y and Yuan H 2021 {\em Sci. China-Phys.
  Mech. Astron.\/} {\bf 64} 107462
  \urlprefix\url{http://engine.scichina.com/doi/10.1007/s11433-021-1747-7}

\bibitem{mu2021swave}
Mu C, Yin Q, Tu Z, Gong C, Lei H, Li Z and Luo J 2021 {\em Chinese Physics
  Letters\/} {\bf 38} 077402
  \urlprefix\url{http://cpl.iphy.ac.cn/EN/abstract/article_105951.shtml}

\bibitem{xu2021multiband}
Xu H~S, Yan Y~J, Yin R, Xia W, Fang S, Chen Z, Li Y, Yang W, Guo Y and Feng D~L
  2021 {\em Phys. Rev. Lett.\/} {\bf 127}(18) 187004
  \urlprefix\url{https://link.aps.org/doi/10.1103/PhysRevLett.127.187004}

\bibitem{xiang2021twofold}
Xiang Y, Li Q, Li Y, Xie W, Yang H, Wang Z, Yao Y and Wen H~H 2021 {\em Nature
  Communications\/} {\bf 12}
  \urlprefix\url{https://doi.org/10.1038/s41467-021-27084-z}

\bibitem{zhao2021nodal}
Zhao C~C, Wang L~S, Xia W, Yin Q~W, Ni J~M, Huang Y~Y, Tu C~P, Tao Z~C, Tu Z~J,
  Gong C~S, Lei H~C, Guo Y~F, Yang X~F and Li S~Y 2021  (\textit{Preprint}
  \eprint{arXiv:2102.08356})

\bibitem{Yu2021}
Yu F~H, Ma D~H, Zhuo W~Z, Liu S~Q, Wen X~K, Lei B, Ying J~J and Chen X~H 2021
  {\em Nature Communications\/} {\bf 12}
  \urlprefix\url{https://doi.org/10.1038/s41467-021-23928-w}

\bibitem{chen2021double}
Chen K, Wang N, Yin Q, Gu Y, Jiang K, Tu Z, Gong C, Uwatoko Y, Sun J, Lei H, Hu
  J and Cheng J~G 2021 {\em Physical Review Letters\/} {\bf 126}
  \urlprefix\url{https://doi.org/10.1103/physrevlett.126.247001}

\bibitem{Du2021}
Du F, Luo S, Ortiz B~R, Chen Y, Duan W, Zhang D, Lu X, Wilson S~D, Song Y and
  Yuan H 2021 {\em Physical Review B\/} {\bf 103}
  \urlprefix\url{https://doi.org/10.1103/physrevb.103.l220504}

\bibitem{ChenCPL}
Chen X, Zhan X, Wang X, Deng J, Liu X~B, Chen X, Guo J~G and Chen X 2021 {\em
  Chinese Physics Letters\/} {\bf 38} 057402
  \urlprefix\url{http://cpl.iphy.ac.cn/EN/abstract/article_105899.shtml}

\bibitem{Zhang2021}
Zhang Z, Chen Z, Zhou Y, Yuan Y, Wang S, Wang J, Yang H, An C, Zhang L, Zhu X,
  Zhou Y, Chen X, Zhou J and Yang Z 2021 {\em Physical Review B\/} {\bf 103}
  \urlprefix\url{https://doi.org/10.1103/physrevb.103.224513}

\bibitem{zhu2021doubledome}
Zhu C~C, Yang X~F, Xia W, Yin Q~W, Wang L~S, Zhao C~C, Dai D~Z, Tu C~P, Song
  B~Q, Tao Z~C, Tu Z~J, Gong C~S, Lei H~C, Guo Y~F and Li S~Y 2021
  (\textit{Preprint} \eprint{arXiv:2104.14487})

\bibitem{yin2021strainsensitive}
Yin L, Zhang D, Chen C, Ye G, Yu F, Ortiz B~R, Luo S, Duan W, Su H, Ying J,
  Wilson S~D, Chen X, Yuan H, Song Y and Lu X 2021 {\em Phys. Rev. B\/} {\bf
  104}(17) 174507
  \urlprefix\url{https://link.aps.org/doi/10.1103/PhysRevB.104.174507}

\bibitem{wang2021competition}
Wang N~N, Chen K~Y, Yin Q~W, Ma Y~N~N, Pan B~Y, Yang X, Ji X~Y, Wu S~L, Shan
  P~F, Xu S~X, Tu Z~J, Gong C~S, Liu G~T, Li G, Uwatoko Y, Dong X~L, Lei H~C,
  Sun J~P and Cheng J~G 2021 {\em Phys. Rev. Research\/} {\bf 3}(4) 043018
  \urlprefix\url{https://link.aps.org/doi/10.1103/PhysRevResearch.3.043018}

\bibitem{Wang2013}
Wang W~S, Li Z~Z, Xiang Y~Y and Wang Q~H 2013 {\em Physical Review B\/} {\bf
  87} \urlprefix\url{https://doi.org/10.1103/physrevb.87.115135}

\bibitem{Isakov2006}
Isakov S~V, Wessel S, Melko R~G, Sengupta K and Kim Y~B 2006 {\em Physical
  Review Letters\/} {\bf 97}
  \urlprefix\url{https://doi.org/10.1103/physrevlett.97.147202}

\bibitem{Guo2009}
Guo H~M and Franz M 2009 {\em Physical Review B\/} {\bf 80}
  \urlprefix\url{https://doi.org/10.1103/physrevb.80.113102}

\bibitem{Kiesel2013}
Kiesel M~L, Platt C and Thomale R 2013 {\em Physical Review Letters\/} {\bf
  110} \urlprefix\url{https://doi.org/10.1103/physrevlett.110.126405}

\bibitem{Wen2010}
Wen J, R\"{u}egg A, Wang C~C~J and Fiete G~A 2010 {\em Physical Review B\/}
  {\bf 82} \urlprefix\url{https://doi.org/10.1103/physrevb.82.075125}

\bibitem{zhang2021firstprinciples}
Zhang J~F, Liu K and Lu Z~Y 2021 {\em Phys. Rev. B\/} {\bf 104}(19) 195130
  \urlprefix\url{https://link.aps.org/doi/10.1103/PhysRevB.104.195130}

\bibitem{tsirlin2021anisotropic}
Tsirlin A~A, Fertey P, Ortiz B~R, Klis B, Merkl V, Dressel M, Wilson S~D and
  Uykur E 2021 {\em arXiv:2105.01397\/} (\textit{Preprint} \eprint{2105.01397})

\bibitem{Lee2021}
Lee S, Collini J, Sun S~X~L, Mitrano M, Guo X, Eckberg C, Paglione J, Fradkin E
  and Abbamonte P 2021 {\em Physical Review Letters\/} {\bf 127}
  \urlprefix\url{https://doi.org/10.1103/physrevlett.127.027602}

\bibitem{WHH}
Werthamer N~R, Helfand E and Hohenberg P~C 1966 {\em Phys. Rev.\/} {\bf 147}(1)
  295--302 \urlprefix\url{https://link.aps.org/doi/10.1103/PhysRev.147.295}

\end{thebibliography}
\end{document}